# Spontaneous magnetic skyrmions in single-layer CrInX$_3$(X=Te, Se)

Wenhui Du, Kaiying Dou, Zhonglin He, Ying Dai*, Baibiao Huang, Yandong Ma*

School of Physics, State Key Laboratory of Crystal Materials, Shandong University, Shandanan Street 27, Jinan 250100, China

*Corresponding author: daiy60@sina.com (Y.D.); yandong.ma@sdu.edu.cn (Y.M.)

**Abstract**

The realization of magnetic skyrmions in nanostructures holds great promise for both fundamental research and device applications. Despite recent progress, intrinsic magnetic skyrmions in two-dimensional lattice are still rarely explored. Here, using first-principles calculations and Monte-Carlo simulations, we report the identification of spontaneous magnetic skyrmions in single-layer CrInX$_3$ (X=Te, Se). Due to the joint effect of broken inversion symmetry and strong spin-orbit coupling, inherent large Dzyaloshinskii–Moriya interaction occurs in both systems, endowing the intriguing Néel-type skyrmions in the absence of magnetic field. By further imposing moderate magnetic field, the skyrmion phase can be obtained and is stable within a wide temperature range. Particularly for single-layer CrInTe$_3$, the size of skyrmions is sub-10 nm and the skyrmion phase can be maintained at an elevated temperature of ~180 K. In addition, the phase diagrams of their topological spin textures under the variation of magnetic parameters of $D$, $J$, and $K$ are mapped out. Our results greatly enrich the research of 2D skyrmionics physics.

**Introduction**

Magnetic skyrmion, a whirling spin texture, has been attracting extensive attention in recent research since it is important for fundamental physics and anticipated to be the next generation of information carriers [1-7]. Each magnetic skyrmion is characterized by a topological invariant called the topological charge Q, which is an integer that cannot be changed by a continuous transformation of magnetism. Because of such a topological nature, magnetic skyrmions are rather robust against environmental disturbance [8,9]. This combined with their small size and efficient current-driven dynamics render magnetic skyrmions promising for novel device applications, such as racetrack memory devices and neuromorphic computing devices [10-12]. Magnetic skyrmions are first experimentally verified in cubic B20 MnSi [13,14] and subsequently in epitaxial magnetic thin films such as Fe/Ir(111) [15] and Ir/Co/Pt [16]. Though highly valuable, such systems suffer from extremely narrow temperature region as well as harsh fabrication technology, limiting their practical application as well as integration [14-18].

Recently, long-range magnetism is discovered in two-dimensional (2D) materials [19-23]. This provides a promising alternative avenue for exploring exotic topologically nontrivial spin phenomena. The physics of magnetic skyrmions usually correlate to the competition between Dzyaloshinskii–Moriya interaction (DMI) and exchange coupling or magnetic anisotropy. Note that most 2D magnetic materials exhibit inversion symmetry, the existence of the essential DMI is excluded in them. There are several strategies to break the inversion symmetry of 2D lattice, such as electric filed [24], proximity effect [25,26], and Janus structure [27-32]. Particularly, Janus structure with intrinsic inversion symmetry breaking has been predicted to be a promising way for establishing significant DMI [33,34]. As Janus MoSSe has been already grown employing different methods [35,36], Janus materials also show high experimental feasibility. Nevertheless, such 2D Janus materials exhibiting magnetic skyrmionics states are rather scarce, and up to now, only a few candidates have been reported [27-32].

Here, via first-principles calculations and Monte-Carlo simulations, we unveil that Janus single-layer (SL) $CrInX_3$ (X=Te, Se), derived from the prototype $In_2X_3$, are promising 2D ferromagnetic semiconductors with skyrmionics physics. Both systems present inherent large DMI, which is originated from the combined effect of broken inversion symmetry and strong spin-orbit coupling (SOC). This yields the intriguing Néel-type skyrmions in both systems in the absence of magnetic field. Upon applying moderate magnetic field, the skyrmion phase can be realized, which is revealed to be robust within a rather wide temperature range. Especially for single-layer $CrInTe_3$, the nontrivial spin texture is sub-10 nm and skyrmion phase can be preserved at an elevated temperature of up to ~180 K, which are technologically desirable. Furthermore, the dependence of their topological spin textures on $D$, $J$, and $K$ are systematically discussed. These results demonstrated that SL $CrInX_3$ can be excellent candidates for skyrmionics applications.

**Methods**

First-principles calculations are performed based on density functional theory (DFT) as implemented in Vienna ab initio simulation package (VASP) [37,38]. The generalized gradient approximation (GGA) in form of Perdew-Burke-Ernzerhof (PBE) functional is used to treat the electron exchange-correlation interactions [39]. The plan-wave cutoff energy is set to 520 eV. The convergence criterion for force and energy are set to 0.01 eV/Å and $1 \times 10^{-6}$ eV, respectively. The vacuum space along z direction is set to 25 Å. The Monkhorst-Pack k-point mesh of $15 \times 15 \times 1$ is adopted to sample the 2D Brillouin zone. $16 \times 4 \times 1$ k-point mesh is adopted for $1 \times 4$ supercell to calculate DMI parameters. To describe well the strong correlations of 3d electrons, GGA +U method is adopted with effective Hubbard U = 3 eV for 3d electrons of Cr atom, as employed in previous works [24,40]. Phonon spectra is obtained by PHONOPY code based on $3 \times 3 \times 1$ supercell [41]. Ab initio molecular dynamics (AIMD) simulations are performed at 300 K for 5 ps with a time step of 1 fs using $4 \times 4 \times 1$ supercell [42].

Parallel tempering Monte-Carlo (MC) simulations [43] with the Metropolis algorithm are carried out using the Hamiltonian of Eq. (3) presented below. All MC simulations are gradually cooled down from an initial disordered state at high temperature (660 K + investigated low temperature) to the investigated low temperature. For each simulated temperature, we used $10^5$ MC steps for thermalization. To obtain the low-energy spin textures of SL CnInX$_3$, the $320 \times 320 \times 1$ supercell with periodic boundary conditions is adopted.

For the convenience of discrete lattice, the topological charge Q is calculated based the method proposed by Berg and Lüscher and following the expression [44]

$$Q = \frac{1}{4\pi} \sum_l qn \quad (1)$$

$$\tan\frac{qn}{2} = \frac{S_i^n \cdot (S_j^n \times S_k^n)}{1 + S_i^n \cdot S_j^n + S_j^n \cdot S_k^n + S_k^n \cdot S_i^n} \quad (2)$$

Here, $S_i^n$, $S_j^n$, $S_k^n$ are the three spin vectors of the n[th] equilateral triangle in the anticlockwise lattice.

**Results and Discussion**

**Figs. 1(a,b)** display the crystal structure of Janus SL CrInX$_3$. It presents the space group of P3m1(C$_{3v}$), and consists five atomic layers stacked in the sequence of X-In-X-Cr-X. Each Cr atom coordinated with six X atoms, forming a distorted octahedral geometry ($l_1 \neq l_2$). Obviously, the inversion symmetry of SL CrInX$_3$ is broken. The lattice constants of SL CrInX$_3$ are listed in **Table S1**. To determine the bonding nature in SL CrInX$_3$, we calculate the electron localized function (ELF). As shown in **Fig. 1(c)** and **Fig. S1**, In-X bond exhibits a covalent character, while Cr-X bond shows an ionic feature. To assess the stability of SL CrInX$_3$, we first calculate its phonon spectra. As shown in **Fig. S2**, there is only a tiny imaginary frequency around the Γ point, indicting its dynamic stability. Its thermal stability is further investigated by performing AIMD simulations. The slight free-energy fluctuation and well-defined structures confirm the thermal stability of SL CrInX$_3$, see **Fig. S3**.

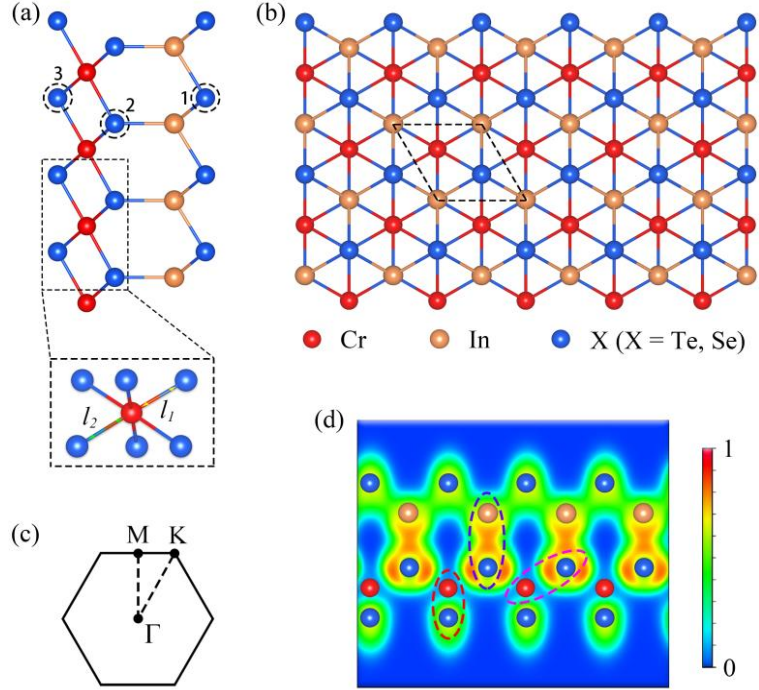

**FIG. 1**. (a) Crystal structure of SL CrInX$_3$ from (a) side and (b) top views. Inset in (a) shows the coordinate environment of Cr atom. In (b), the unit cell is represented by the black dotted lines. (c) 2D Brillouin zone. (d) Map of electron localization function (ELF) of SL CrInTe$_3$, where 0 and 1 indicate vanished and accumulated electron densities, respectively.

Concerning Cr atom, its valence electronic configuration is 3d$^4$4s$^2$. In SL CrInX$_3$, each Cr atom donates three electrons to the six coordinated X atoms, resulting in the valence electronic configuration of 3d$^3$4s$^0$. Under the distorted octahedral crystal field, Cr-3d orbitals split into two groups: higher-lying e$_g$ ($d_{xy}, d_{x^2-y^2}$) and lower-lying t$_{2g}$ ($d_{xz}, d_{yz}, d_{z^2}$). The three left electrons of Cr atom would half-occupy t$_{2g}$ orbitals, yielding a magnetic moment of 3 μ$_B$ on each Cr atom. As shown in **Fig. 2(a)**, the magnetic moment is mainly distributed on the Cr atoms.

To investigate the interactions of magnetic moments in SL CrInX$_3$, we adopt a Heisenberg spin Hamiltonian:

$$H = -J \sum_{\langle i,j \rangle} \mathbf{S}_i \cdot \mathbf{S}_j - \lambda \sum_{\langle i,j \rangle} S_i^z S_j^z - K \sum_i (S_i^z)^2 - \sum_{\langle i,j \rangle} \mathbf{D}_{ij} \cdot (\mathbf{S}_i \times \mathbf{S}_j) - mB \sum_i S_i^z \quad (3)$$

Here, $\mathbf{S}_i$ is the unit vector of Cr atom at site $i$. $\langle i,j \rangle$ represents the nearest-neighbor sites. $J$, $\lambda$, $K$ and $D_{ij}$ represent the parameters of Heisenberg isotropic exchange, anisotropic symmetric exchange, single ion anisotropy and DMI, respectively. The last term is Zeeman energy, where $m$ and $B$ stand for the on-site magnetic moment of Cr atom and the external magnetic field, respectively. The magnetic parameters of $J$, $\lambda$, and $K$ are obtained by considering four different magnetic configurations, see **Fig. S4**. As listed in **Table S1**, $J$ are calculated to be 26.47 and 26.36 meV, respectively, for SL CrInSe$_3$ and CrInTe$_3$. The positive values indicate that the magnetic interactions in both systems prefer ferromagnetic (FM) coupling. Such FM coupling is related to their structures, namely, the Cr-X-Cr bonding angle is close to 90° for both systems, wherein FM super-exchange would dominate the magnetic exchange interaction according to the Goodenough-Kanamori-Anderson rules [45,46]. **Figs. S5** displays the band structures of SL CrInSe$_3$ and CrInTe$_3$, which are semiconductors with an indirect bandgap of 1.05 and 0.25 eV, respectively. Therefore, both systems are 2D FM semiconductors.

For stablishing long-range FM coupling, magnetic anisotropy is essential. As listed in **Table S1**, SL CrInTe$_3$ exhibits $K = 1.544$ meV and $\lambda = 0.1003$ meV, suggesting large out-of-plane magnetic anisotropy. Different from SL CrInTe$_3$, SL CrInSe$_3$ shows $K = 0.262$ meV and $\lambda = -0.0063$ meV, indicating that its single ion anisotropy and anisotropic symmetric exchange prefer in-plane and out-of-plane magnetizations, respectively. Given the fact that K is two orders of magnitude larger than $\lambda$, SL CrInSe$_3$ also favors out-of-plane magnetization. It is interesting to note that although the $\lambda$ and $K$ parameters in these two systems are significantly different, their $J$ parameters are roughly identical, as shown in **Fig. 2(b)**. This discrepancy is sought into the difference in SOC strength, which is proportional to $Z^4$ (Z is the atomic number). Compared with Se atom, Te atom is heavier, so the SOC effect in CrInTe$_3$ is stronger, resulting in more significant anisotropic symmetric exchange $\lambda$ and single ion anisotropy $K$. In contrast, for Heisenberg isotropic exchange $J$, it is less affected by SOC. Based on the magnetic parameters of $J$, $\lambda$, and $K$, the Curie temperature $T_c$ of SL CrInTe$_3$ and CrInSe$_3$ is estimated to be 340 K and 295 K, respectively, which is dramatically lager than that of CrI$_3$ (45 K) [19] and bilayer Cr$_2$Ge$_2$Te$_6$ (28 K) [21].

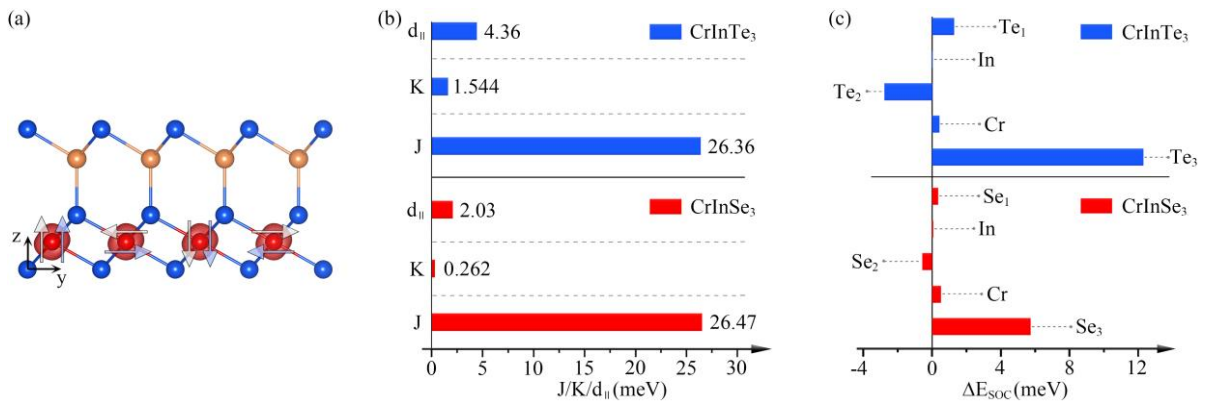

**FIG. 2**. (a) Spin charge density of SL CrInTe$_3$. White/blue vectors in (a) indicate the left/right-hand spin-spiral configurations used to obtain the in-plane DMI parameters $d_\parallel$ and atomic projected SOC energy. (b) Isotropic exchange coupling parameter $J$, single ion anisotropy parameter $K$ and in-plane DMI parameter $d_\parallel$ of SL CrInX$_3$. (c) SOC energy projected on atoms ($\Delta E_{soc}$) for SL CrInX$_3$.

We then consider the left- and right-hand spin-spiral configurations [**Fig. 2(a)**] to obtain the DMI parameters $\boldsymbol{D}_{ij}$. According to Moriya's rule [34], because the mirror plane passes through the middle of the bond between two adjacent Cr atoms, the DMI vector $\boldsymbol{D}_{ij}$ for the nearest-neighboring Cr atoms is perpendicular to their bond. $\boldsymbol{D}_{ij}$ can be written as $\boldsymbol{D}_{ij} = d_\parallel(\boldsymbol{z} \times \boldsymbol{u}_{ij}) + d_\perp \boldsymbol{z}$, where $\boldsymbol{z}$ and $\boldsymbol{u}_{ij}$ represent the unit vectors pointing along the z direction and from site i to site j, respectively. Since the out-of-plane component of $\boldsymbol{D}_{ij}$ plays a negligible role in the formation of skyrmions, we only focus on its in-plane component $d_\parallel$. Remarkably, $d_\parallel$ in SL CrInTe$_3$ (CrInSe$_3$) is rather large, which is calculated to be 4.36 (2.03) meV. Especially for SL CrInTe$_3$, $d_\parallel$ is much larger than the values reported in most previous studies [15,29-32,47]. More importantly, the $d_\parallel/J$ ratio is found to be 0.165 (0.077) for SL CrInTe$_3$ (CrInSe$_3$), which is within (near) the typical range of 0.1- 0.2 for more likely forming magnetic skyrmions [48].

To get further insight into the DMI in SL CrInX$_3$, we project the associated SOC energy $\Delta E_{soc}$ [$\Delta E_{soc}$ is defined as the energy difference between left- and right-hand spin-spiral configurations] on each atom. As shown in **Fig. 2(c)**, the main DMI contribution does not originate from the magnetic Cr atoms, but from its adjacent X atoms. In this case, the X atoms, serving as effective sites of SOC, introduce the spin-orbit scattering necessary for DMI. This clearly indicates that the Fert-Levy mechanism [49,50] is responsible for DMI in SL CrInX$_3$. As larger Z corresponds to stronger SOC, more significant DMI is obtained in SL CrInTe$_3$. In addition, we find that $|\Delta E_{soc}|$ from X$_3$ atom is much larger than that from X$_2$ atom. This correlate to the strong covalent bonding between X$_2$ and In atoms, which weakens the coupling between X$_2$ and Cr atoms and thus results in a relatively small DMI contribution from X$_2$ atom.

Based on the first-principles parametrized Hamiltonian of Eq. (3), parallel tempering MC simulations are performed to explore the possible topological spin textures in SL CrInX$_3$. The topological stability of magnetic skyrmions is characterized by topological charge Q, which is defined as $Q = \frac{1}{4\pi} \int \boldsymbol{m} \cdot \left(\frac{\partial \boldsymbol{m}}{\partial x} \times \frac{\partial \boldsymbol{m}}{\partial y}\right) dx dy$ [9,44]. Here, $\boldsymbol{m}$ is the normalized magnetization vector. **Figs. 3(a,b)** present the spin textures of SL CrInX$_3$. Without applying magnetic field, labyrinth domains are observed in both systems at 0 K. Aside from labyrinth domains, intriguingly, the Néel-type magnetic skyrmions with $Q = \pm 1$ also exist in both systems spontaneously. Such a mixed phase is referred to as spin spiral phase (SS) below. It is worthy emphasizing that for SL CrInTe$_3$, the magnetic skyrmions occur in different domains, which presents opposite polarization and thus opposite topological charge Q; see **Fig. 3(b)**. Moreover, the diameter of magnetic skyrmions in CrInTe$_3$ is about 7.6 nm, significantly smaller than those of MnSTe (40 nm) [28], LaCl/In$_2$Se$_3$ (23 nm) [25] and Cr(I, Cl)$_3$ (10.5 nm) [29]. Such a small size within sub-10nm is technologically desired for future skyrmionics devices. Different from the case of CrInTe$_3$, the magnetic skyrmions in SL CrInSe$_3$ occur in one domain, and its size is about 27.8 nm.

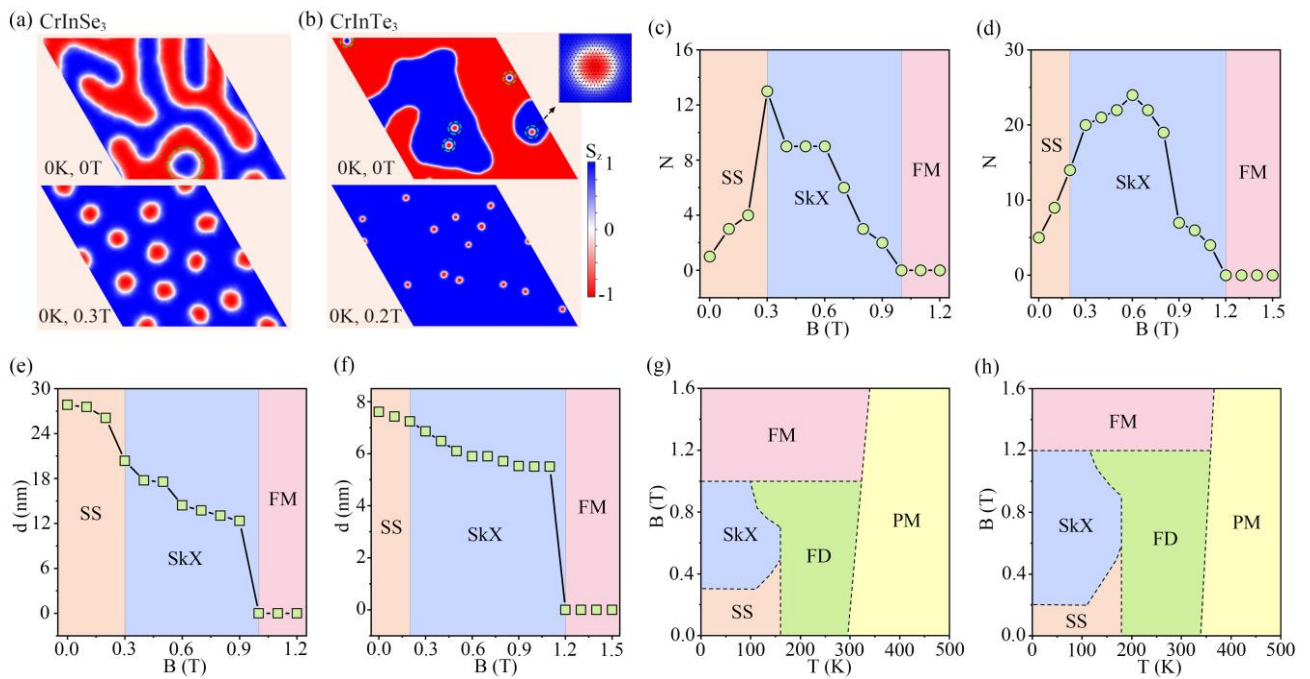

**FIG. 3**. Spin textures of SL (a) CrInSe$_3$ and (b) CrInTe$_3$ without and with applying magnetic field (B) at 0 K. Inset in (b) shows the magnetization distribution in skyrmion. Colors are assigned for out-of-plane magnetization components and black arrows represent the in-plane components. Number (N) of skyrmions for SL (c) CrInSe$_3$ and (d) CrInTe$_3$ as a function of B. Skyrmion diameters for SL (e) CrInSe$_3$ and (f) CrInTe$_3$ as a function of B. Magnetic phase diagrams for SL (g) CrInSe$_3$ and (h) CrInTe$_3$ with temperatures T and B. SkX, SS, FM in (b-h) and FD, PM in (g, h) indicate skyrmion, spin spiral, fluctuation-disorder, ferromagnetic and paramagnetic phases, respectively.

We then explore the effect of external magnetic field on the topological spin structures of SL CrInX$_3$. For SL CrInTe$_3$, the labyrinth domain shrinks with increasing magnetic field, and disappears completely at 0.2 T, see **Fig. 3(b)**. In this case, only isolated skyrmions are left in a FM background, forming the magnetic skyrmion (SkX) phase. Importantly, as shown in **Fig. 3(d)**, this SkX phase can be preserved in a wide magnetic field range of 0.2 – 1.2 T. For the density of skyrmions, it increases with increasing the magnetic field from 0 T to 0.6 T. For the cases of 0.6 – 1.2 T, the density of skyrmions decreases significantly, and the skyrmions pattern is completely magnetized into the trivial FM phase at 1.2 T. **Fig. 3(f)** shows the impact of magnetic field on the skyrmion size for SL CrInTe$_3$, from which we can see that the skyrmion size decreases with increasing magnetic field. In detail, the skyrmion diameter reduces from about 7.6 nm at 0 T to 5.5 nm at 1.1 T and then shrinks into a FM background at 1.2 T. The underlying physics for this phenomenon is related to that fact that external magnetic field favors the out-of-plane magnetization as that of the FM background.

The roughly similar scenario is also observed in SL CrInSe$_3$. With increasing magnetic field, the spin structure of SL CrInSe$_3$ undergoes a SS-SkX-FM phase transition. All the labyrinth domains shrink at 0.3 T, leading to the skyrmion lattice, see **Fig. 3(a)**. By further increasing the magnetic field, the density of skyrmions decreases, and drops to zero at 1 T, which gives rise to a trivial FM phase. With respect to the case of SL CrInTe$_3$, the SkX phase of SL CrInSe$_3$ can be preserved in a relatively narrow magnetic field range of 0.3 – 1 T [**Fig. 3(c)**]. Moreover, the increase of magnetic field also results in the shrinking of skyrmion size before transforming into FM phase, i.e., the skyrmion size decreases from 27.8 nm at 0 T to 12.3 nm at 0.9 T, see **Fig. 3(e)**.

Considering that thermal fluctuations may destabilize magnetic order, we further study the temperature effect on topological spin structures in SL CrInX$_3$ under various magnetic fields. The corresponding magnetic phase diagrams for SL CrInX$_3$ are displayed in **Figs. 3(g,h)**. We can see that besides SS, SkX and FM phases, two new phases [i.e., fluctuation-disorder (FD) and paramagnetic (PM) phases] emerge. In FD phase, arising from thermal fluctuation, on one hand, the vortex of skyrmion is partially deformed, and on the other hand, the breaking of skyrmions increases the number of little vortices. The former and latter factors would decrease and increase topological charge Q. Here, we take $|Q − 1| > 0.01$ as the judgment standard for the transformation of SkX phase into FD phase. In PM phase, magnetic order is completely destroyed, and thus spin texture becomes disordered as well. The Curie temperature under different external magnetic fields is calculated to judge the critical temperature of phase transition into PM phase.

From **Fig. 3(h)**, we can see that SL CrInTe$_3$ is stabilized in SS phase at low temperature and low magnetic field (B < 0.2 T). With increasing temperature, the SS phase is transformed into FD and PM phases, successively. When external magnetic field approaches 0.2 T, the labyrinth domain completely shrinks to isolated skyrmions at 0 K, suggesting the transition from SS to SkX phases. Under the magnetic field of 0.2 – 0.6 T, with rising temperature, the SkX phase is transformed into SS phase, then FD phase and finally PM phase. While under 0.6 – 1.2 T, with increasing temperature, the SkX phase is directly transformed into FD phase, and then PM phase. It is worth noting that when the magnetic field exceeds 0.9 T, the maximum temperature that the skyrmions can withstand gradually decreases. This suggests that moderate magnetic field can be used to stablish skyrmions. Under magnetic field above 1.2 T, the SkX phase is magnetized into FM phase. This scenario is roughly shared by the case of SL $CrInSe_3$; see **Fig. 3(g)**. Therefore, by modulating external magnetic field and temperature, we can control the generation and annihilation of skyrmions, and thus the topological phase transition, in SL $CrInX_3$. We wish to point out that the skyrmions of SL $CrInX_3$ can be obtained in a considerable wide temperature range. Especially for SL $CrInTe_3$, the SkX phase can be preserved as temperatures rising up to ~180 K.

To deeply understand the skyrmionics physics in SL $CrInX_3$, taking SL $CrInTe_3$ as an example, we investigate the relationships between topological spin structures and magnetic parameters of $D$, $J$ and $K$. Since $D/J$ is regarded as an important parameter to estimate the existence of skyrmions, we first explore the dependence of spin texture on $D/J$ under different external magnetic fields, while keeping other parameters the same as those in SL $CrInTe_3$ expect $\lambda$. Considering $\lambda$ in SL $CrInX_3$ is negligible, it is set to zero in the simulation for simple. As shown in **Fig. 4(a)**, the spin texture undergoes a FM-SkX-SS transition with increasing $D/J$ under different external magnetic fields. In SkX region, the size and density of skyrmions increase with increasing $D/J$. In SS region, as the $D/J$ increases, the density of labyrinth domains increases, while the size reduces. Moreover, as shown in **Fig. 4(a)**, when the magnetic field increases, the SkX phase moves to larger $D/J$ region. This indicates that a larger external magnetic field is required for stabilizing the skyrmions in system with larger $D/J$.

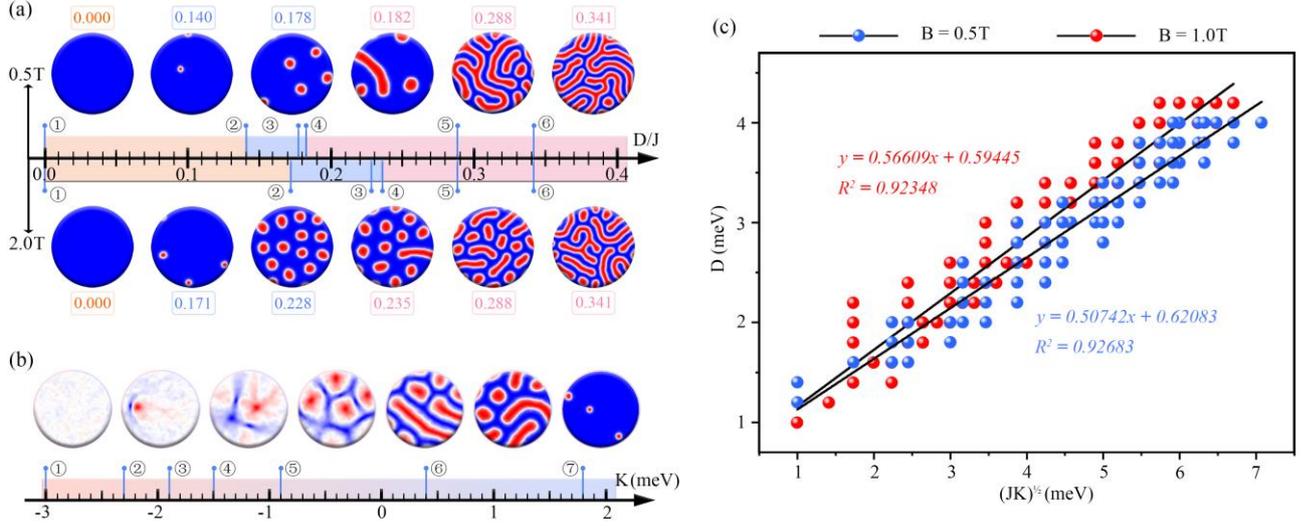

**FIG. 4**. (a) Evaluation of spin textures as a function of $D/J$ under different external magnetic fields. (b) Evaluation of spin texture as a function of $K$ under the external magnetic field of 0.5 T. (c) Linear scaling relation between $D$ and $\sqrt{JK}$ of SkX phase under different external magnetic fields.

The evaluation of spin textures as a function of $D/J$ under different magnetic fields is related to the competition among $D$, $J$ and external magnetic field $B$. Specifically, $J$ and $B$ tend to force the parallel arrangement of magnetic moments, while $D$ induces the magnetic moments to be arranged noncollinearly. In the SkX region, when $D/J$ is relatively small, the magnetic exchange coupling plays a more important role, which reduce the vortices in the FM background and make the magnetic moments' rotation from the **-z** direction at the center of skyrmion to **z** direction at the boundary at a small atomic scale. In this case, the skyrmions with relatively small-size and low-density can be obtained. Upon decreasing $D/J$ close to zero, DMI is negligible, and the strong magnetic exchange coupling $J$ leads the parallel arrangement of magnetic moments, i.e., FM phase. On the other hand, when $D/J$ is large enough, DMI plays a dominated role, benefiting for the emergence of labyrinth domains. With these results in hand, we can understand the evaluation of spin textures discussed above.

After establishing the roles of $D$ and $J$, also taking SL CrInTe$_3$ as an example, we next study the effect of $K$ on the topological spin texture. As shown in **Fig. 4(b)**, with increasing $K$ from -3 to 2 meV, the spin texture undergoes an iFM (in-plane FM)-bimeron-SS-SkX phase transition. Moreover, we find that the size of topological spin textures decreases with increasing $|K|$. When $K$ is positive, the perpendicular magnetic anisotropy provides the energy gain to sustain an out-of-plane spin alignment, thus promoting the formation of Néel-type skyrmions. For $K < 0$, the system favors in-plane spin alignment, converting the Néel-type skyrmions into bimerons. Furthermore, similar to $J$, $K$ is prone to force the magnetic moment in a parallel arrangement. In this regard, a large $|K|$ can suppress the modulated labyrinth domains and skyrmions, and the reduction of the size of skyrmions or bimerons.

From above, we can see that $D$, $J$ and $K$ can significantly affect the formation and characters of SkX phase. To help the rational design of SkX phases, it is important to identify a universal descriptor based on these three magnetic parameters. To this end, we conduct high throughput calculations with changing each magnetic parameter independently under different magnetic fields. As shown in **Fig. 4(c)**, intriguingly, relationships between $D$ and $\sqrt{JK}$ of SkX phase are linear under different magnetic fields, and a stronger magnetic field corresponds to a larger slope. This suggests that $D/\sqrt{JK}$ can be used to estimate the required external magnetic field for forming SkX phase.

**Conclusion**

To summarize, we propose that SL CrInX$_3$ are compelling 2D ferromagnetic semiconductors with skyrmionics physics on the basis of first-principles calculations and Monte-Carlo simulations. We find that, due to the inherent large DMI, both systems exhibit Néel-type skyrmions, without needing the external magnetic field. Moreover, the skyrmion phase can be established in both systems under a moderate magnetic field, and such phase is stable within a rather wide temperature range. In particular for SL CrInTe$_3$, the skyrmion size is sub-10 nm, and SkX phase can be preserved at an elevated temperature of up to ~180 K. In addition, we also systematically investigate the dependence of the topological spin textures on $D$, $J$, and $K$, and reveal that $D/\sqrt{JK}$ can be used to estimate the required magnetic field for forming SkX phase.

**Acknowledgement**

This work is supported by the National Natural Science Foundation of China (No. 12074217), Shandong Provincial Natural Science Foundation (Nos. ZR2019QA011 and ZR2019MEM013), Shandong Provincial Key Research and Development Program (Major Scientific and Technological Innovation Project) (No. 2019JZZY010302), Shandong Provincial Key Research and Development Program (No. 2019RKE27004), Shandong Provincial Science Foundation for Excellent Young Scholars (No. ZR2020YQ04), Qilu Young Scholar Program of Shandong University, and Taishan Scholar Program of Shandong Province.

**Competing interests**

The authors declare no competing interests.

**Date availability**

The authors declare that the data supporting the findings of this study are available within the paper and its supplementary information files.